# Single femtosecond laser pulse interaction with mica

Saurabh Awasthi[1*], Douglas J. Little[1], A. Fuerbach[1], D.M. Kane[1*]

1. MQ Photonics research centre, Department of Physics and Astronomy, Macquarie University, Sydney, NSW-2109

**Ultrafast, femtosecond laser pulse interaction with dielectric materials has shown them to have significantly higher laser fluence threshold requirements, as compared to metals and semiconductors, for laser material modification, such as laser ablation. Examples of dielectrics are crystalline materials such as quartz and sapphire, and amorphous glasses. The interaction between femtosecond laser pulses, at a wavelength with negligible linear absorption, and a dielectric has been found to be weak, and multiple pulse irradiation is therefore typically used in order to see significant and quantifiable effects. In this study the dielectric is the crystalline, layered, natural mineral muscovite, a mica with formula $KAl_2(Si_3Al)O_{10}(OH)_2$. Muscovite, newly cleaved, is used in a wide range of technological and scientific applications including as an insulating material in electronics and as an ultra-flat and ultra-clean substrate. A single, ~800 nm wavelength, ~6 micron spotsize, ~150 fs laser pulse is found to lead to a systematic range of laser modification topologies, as a function of the fluence of the single laser pulse, including bulk removal of material. The fs laser pulse/material interaction is greater than expected for a standard dielectric at a given fluence. Optical surface profiling and FESEM are used to characterise the topologies. Contrasting the results of the two techniques supports the use of optical surface profiling to characterise the material modification despite its limitations in lateral resolution as compared to FESEM. The interlayer mineral water content of natural muscovite is proposed as the primary reason that mica behaves differently to a standard dielectric when irradiated with a single 800 nm fs laser pulse.**

## Introduction

The interaction of ultrafast femtosecond (fs) laser pulses with dielectric materials[1,2,3], metals[4,5] and semiconductors[6,7] has been extensively studied, including theoretical modeling[8,9]. In metals, free charge absorption, leads to transfer of the energy from the fs laser pulse to the bulk[10,11]. Direct absorption of the fs laser pulses, leads to low experimentally observed laser ablation thresholds for metals: ~ 0.4-0.9 J/cm$^2$ [12,13]. In contrast, in transparent dielectrics, the pulse energy is non-linearly absorbed in the material. First, a critical density of free electrons is generated by the incident fs laser pulse to create defect states in the system. Thereafter, the rest of the pulse energy is absorbed by the generated electron cloud and transferred to the bulk[3,14–16]. This nonlinear ionization, leads to a higher ablation threshold of ~2.5-15.2 J/cm$^2$ [17–19] in dielectrics using multiple laser pulse irradiation. Hence, in large bandgap dielectric materials, multiple pulses have been used in studies measuring the ablation threshold and seeking to remove bulk material by laser ablation[18,19]. On longer time scales, beyond the pulse duration, in metals, the energy is transferred to the bulk by electron-lattice heating followed by heterogeneous or homogenous melting leading to ablation in the processed region[8]. In semiconductors and dielectrics, a plasma state or non-thermal melting follows the free electron generation and leads to ablation[3,20,21].

Here, we show that non-linear, multiphoton interaction of a single femtosecond laser pulse with muscovite sheet leads to surface modification at a *single laser pulse threshold*, which is found to be unprecedentedly low for a dielectric. Topologies of the processed regions as modified by the pulse, are unconventional compared to any observed published study of femtosecond laser pulse interaction with a dielectric. We propose that the novelty of the results derives from the unique layered structure of muscovite and mineral water trapped between the layers. Our study observes modified topologies on the mica surface from "*bumps*" to "*craters*" with variation in single pulse fluence. Bumps are most likely due to localized de-adhesion of layers driven by the mineral water in the laser affected region.

The abundant natural mineral muscovite has a unique layered structure and composition leading to its distinct physical and dielectric properties [22,23,24] that are used in a range of technological applications. Layered muscovite has a very low cleavage energy of ~ 500 mJ/m$^2$ [25,26], along the preferred plane of delamination, the (100) basal plane. This ability of muscovite to be easily cleaved along this plane is unique among the naturally occurring minerals. Such delamination commonly yields an atomically flat cleaved surface over a large surface area [27,28]. This makes it a substrate of choice for various depth sensitive (height) characterization techniques, self-assembly[29,30], and thin film growth[31–33]. Experimental studies in the past have measured the adhesion energy [34,35] between layers

*Corresponding authors at: MQ Photonics research centre, Department of Physics and Astronomy, Macquarie University, Sydney, NSW-2109.

E-mail addresses: saurabh.awasthi@hdr.mq.edu.au ; deb.kane@mq.edu.au

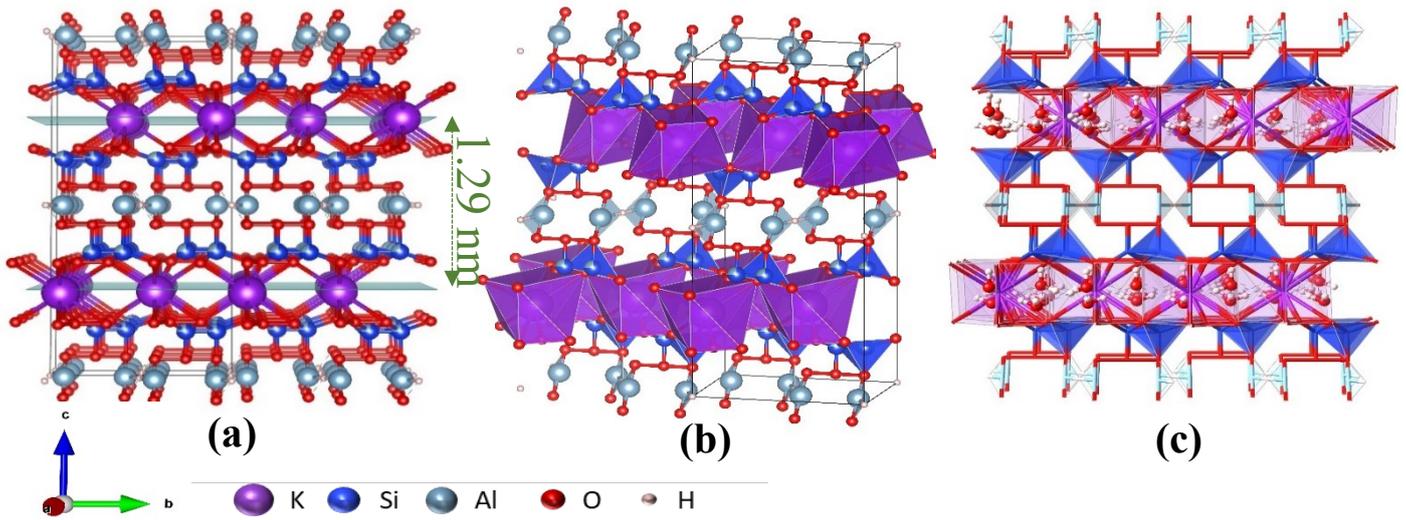

**Figure 1:** Crystal structural configuration of muscovite. a) Ball and stick representation of the structure. Green planes represent the basal plane (001), the thickness of an individual layer is 1.29 nm. b) Polyhedral depiction showing, $AlO_6$ octahedral layer sandwiched between two $SiO_4$ tetrahedral layers. c) Ball and stick representation highlighting the $H_2O$ molecules, other atoms in the structure are made invisible and octahedral layer is transparent. Orientation of the planes are as shown in the vector notation. Purple balls represent potassium, blue- Silicon, grey- aluminium, red- oxygen and gold-hydrogen. Bi-colour lines show the bonding among atoms.

and the geometry of cracks and blisters[36] associated with the cleavage of planes or delamination.

Muscovite is a mica and a phyllosilicate with sum formula[37] $KAl_2(Si_3Al)O_{10}(OH)_2$. Phyllosilicates are the sheet silicates, consisting of a central Si atom surrounded by four oxygen atoms at the corner of tetrahedron. The structural composition of muscovite is visualized in *fig.1*. An octahedral layer of $AlO_6$ is sandwiched between two tetrahedral layers of $SiO_4$, sometimes referred to as di-octahedral mica. The tetrahedral sheets form a honeycomb network with basal oxygen linking the adjacent tetrahedrons. Aluminium (Al) atoms partially substitute Si ions to give the (001) basal layer a net negative charge. This net negative surface charge at the layer boundary is compensated by the interlayer cation $K^+$ and the layers are held together by these compensatory interlayer cations. The thickness of an individual mica monolayer is computed to be ~ 1.29nm[38], *shown in fig.1(a)*.

Water in muscovite

Affinity of water to mica was first documented by Langmuir in 1918[39]. Mica was hypothesized to cleave along layers enclosing *mineral water* molecules (*fig.1(c)*) resulting in a thin primary water film being intimately absorbed at the top cleaved layer. The adsorbed condensed water in muscovite is stored in a capillary like configuration and regarded to be a derivative of water[35] with a higher refractive index and lower vapour pressure compared to pure water. Sakuma analysed the optimum sites for water molecule accumulation within the muscovite lattice[34] and found that between 2 and 12 $H_2O$ molecules can surround the single muscovite unit cell.

Significant experimental effort has been put into identifying the physical properties of muscovite. In thermal studies, muscovite was kept at a constant temperature of ~800°C for 760 hours[40], resulting in de-hydroxylation – the removal of hydroxyls, in the form of water, from the crystal structure. Two hydroxyl ions in muscovite condense to form one $H_2O$ molecule. In the de-hydroxylation process, escaping gas pockets, that arise from hot $H_2O$ molecules, were speculated to be responsible for the observed delamination of the muscovite layers [41,42].

X-Ray Fluorescence (XRF) elemental analysis of the 300μm thick muscovite sheets used in this study, estimates a 4.72 weight percent (%) loss on ignition (L.O.I), as shown in *fig.2*. XRF is reported as oxides, rather than as aluminosilicates that the majority of aluminium and silicon in mica occurs in. The computation of each cation and its oxide percentage from the weight percentage data of XRF implies that all the O cations in the crystal are accounted for. Hence, we estimate the L.O.I as the weight percentage of mineral water present in the muscovite sheet. These estimates are close to the reported data of $H_2O$ in muscovite[37]. Factoring in the 4.72 weight percentage of water in muscovite, leads to the estimate of ~ 4 ± 1 $H_2O$ molecules per unit cell[34] (Supplementary section D). The volume of mica affected by a laser pulse with fluence in the range of 2.47 - 7.07 J/cm$^2$ encapsulates $2.44*10^7$ - $9.65*10^9$ $H_2O$ molecules (Supplementary section E).

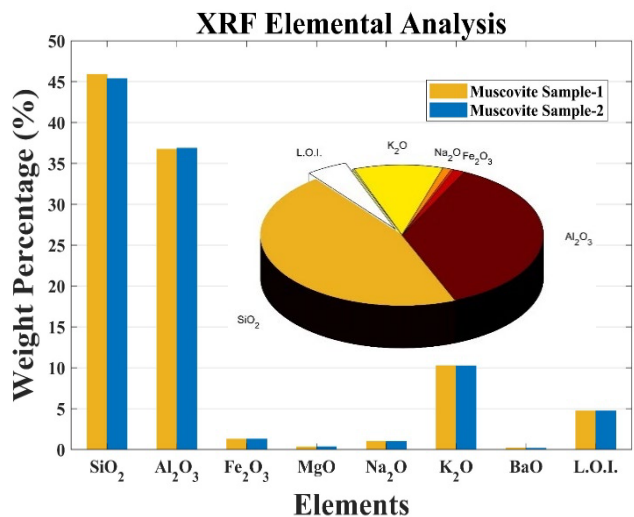

**Figure 2:** Elemental analysis using x-ray fluorescence, vertical axis enlists the detected elements against the weight percentage at the horizontal axis. Two muscovite samples from same sheet depicted in blue and yellow, were put through the test to get the statistical validation of the data. L.O.I. shows the loss on ignition. We predict it to be $H_2O$, being the most probable candidate and close to previous estimates[37].

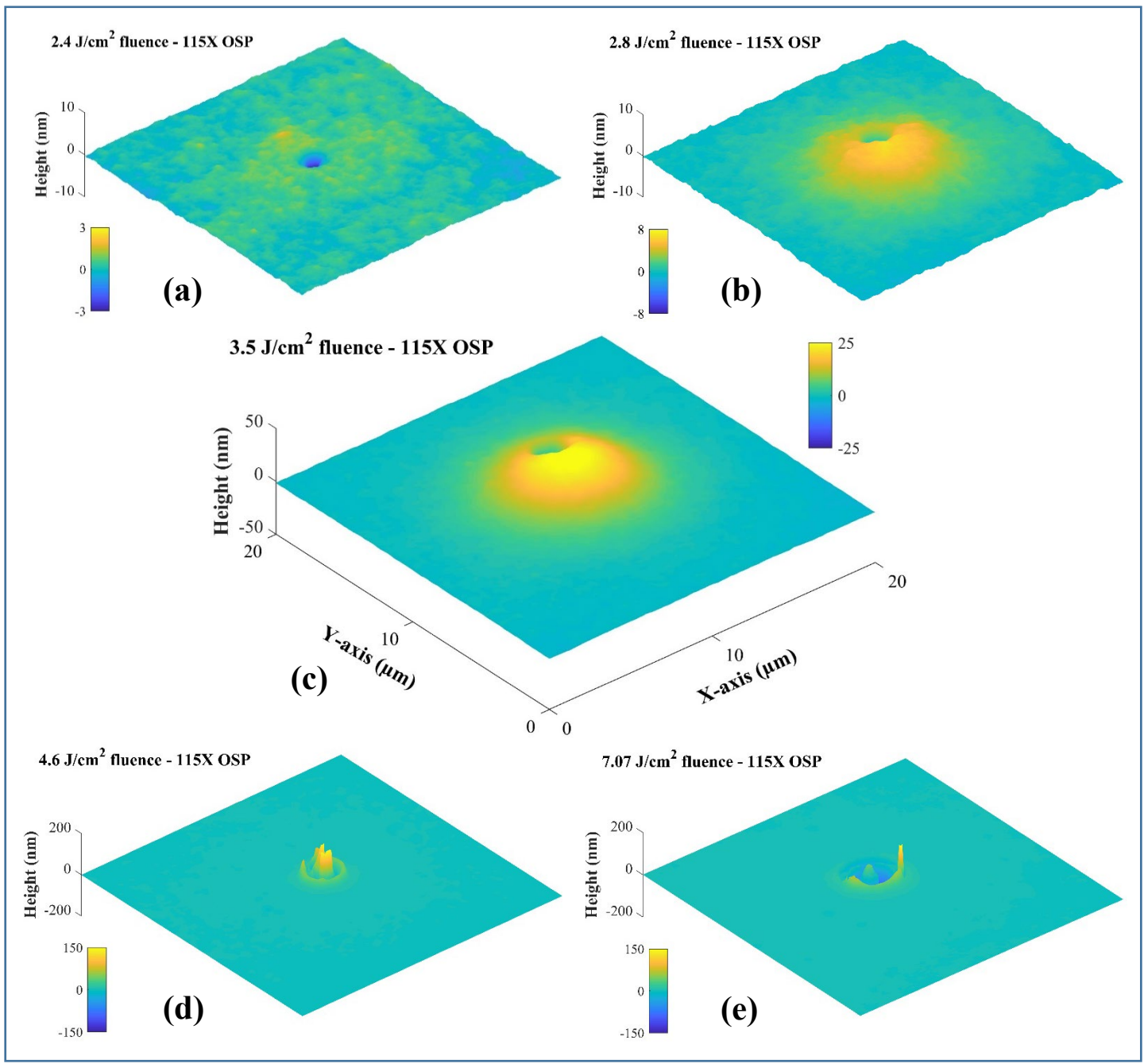

**Figure 3:** Three-dimensional (3-D) Optical surface profiles micrograph of the single femtosecond (fs) laser pulse processed sites. a) Fluence 2.4 J/cm$^2$: Crater with depth 3.8nm and diameter 1.51µm. b) Fluence 2.8 J/cm$^2$: bump is 6.75 nm high and affected diameter of 7.85µm. Crater within the bump region is 10.04nm deep and 1.31µm diameter. c) Fluence 3.5J/cm$^2$, higher surface bump of 26.60nm and diameter 13.02µm. d) Deep crater (below surface) with rim (above surface) surrounding it and high central jet (above surface) at fluence 4.6 J/cm$^2$. Crater is 141.8 nm deep, rim is 58.29 nm high with diameter 2.49µm and central jet is 137.44 nm above surface. e) At high fluence of 7.07 J/cm$^2$, crater deepens to 138.72 nm, rim is 31.37nm high with diameter 4.52µm, central jet confines below the surface. X and Y axis calibration shown in figure (c) applies to all of (a)-(e). Colour bar alongside each map represents the height along Z axis with reference to surface at zero.

<u>Laser pulse modified surface topology</u>

The topology of the modified region was characterized using optical surface profilometry (OSP), to identify the geometrical parameters of the processed region, and by field emission scanning electron microscopy (FESEM) to confirm the structure. Topologies for laser processed sites, as a function of single laser pulse fluence, as measured by OSP are shown in fig. 3. These have sub nanometre height resolution and ~0.3 µm lateral resolution. Height and width data for key features from the topologies are summarized in fig 4.

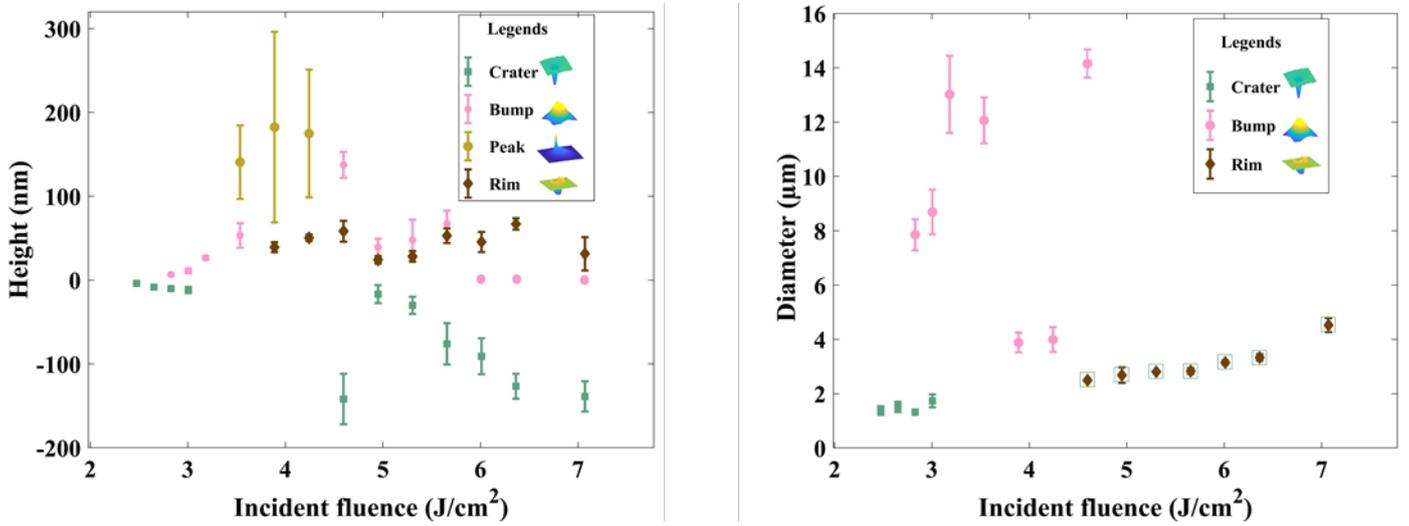

**Figure 4:** Plot exhibiting the geometrical measurements of the fs pulse processed sites at fluences in the range 2.4 J/cm$^2$-7.07 J/cm$^2$. a) Height of varied surface features as measured with reference to surface basal plane, positive and negative values show the modification above and below the surface respectively. b) Diameter of the affected region by the pulse. Legends in the inset exhibit markers for each distinct surface modification, square (green)-craters, small circle (pink)-bump, big circle (olive green)- jet and diamond (brown)-rim. Green box overlapping the brown diamond shows, the crater diameter is approximately same as the rim diameter. The horizontal bar across each marker show the standard deviation in the measurement, at least 20 sites were measured to derive its value. Height of the bumps at zero level in (a), indicate bumps are located inside the deep craters. Their height is either equal to the depth of the crater or less.

At a fluence of 2.4 J/cm$^2$, a shallow crater with a depth (3.84 ± 1.07) nm and a diameter (1.37 ± 0.16) µm is obtained as shown in *fig. 3(a)*. No surface modifications are observed by any of the characterization techniques used, at fluences lower than 2.4 J/cm$^2$. 25 craters are measured to identify the average depths and diameters. Hence, this fluence is the threshold for single fs laser pulse modification of muscovite. Thus, the first topology is a *"crater"*, with a negative height. At a fluence of 2.8 J/cm$^2$, the processed sites extend both above the original surface, a *"Bump"*, and below, a *"crater"* as shown in *fig. 3(b)*. The crater deepens to (10.04 ± 1.66) nm and (1.31 ± 0.09) µm wide. However, the crater is now located within a bump of height (6.75 ± 0.89) nm above the surface.

*Bumps* become more apparent at a fluence of 3.5 J/cm$^2$ as *shown in fig. 3(c)*, 1.46 times the threshold fluence for measurable modification. The average height of the detected bumps increases to (26.60 ± 2.06) nm. At this fluence, the modification is limited to being entirely above the surface. Bumps in the fluence region below the bump threshold, 2.8-3.5 J/cm$^2$, surround a non-central crater. Also, the diameter of the affected region i.e. (7.85 ± 0.56) µm – (13.02 ± 1.42) µm, in this fluence region, is larger than the spot size of the laser beam, 6 µm. A further increase in the fluence to at least 1.9 times the *crater* threshold, i.e. 4.6 J/cm$^2$, introduces a new feature: A rim that is surrounding the popped-up, bump topology, as shown in *fig. 3(d)*. Finally, at the highest fluence employed in this study, 7.07 J/cm$^2$, the height of the bumps is observed to reduce (*fig.3e*), above the surface. However, the craters deepen, and the bumps are situated inside those craters. This, in effect, is still an increase in the absolute height of the bumps compared to previous fluences. The progression of the modified surface topologies as a function of incident single pulse fluence is depicted in *fig.4*. *Fig 4a* shows the height of key features of a laser modified topology at a given fluence and *fig. 4b* shows the diameter of key features. After the modification threshold, the craters (negative height given in *fig.4a* and diameter given in

*fig.4b*), grow until the onset of the appearance of a bump at a fluence ~2.8 J/cm$^2$. In the fluence range 2.8 - 3.0 J/cm$^2$ the height of the bump and the depth of the crater grow systematically. In the fluence range 3.0 - 3.53 J/cm$^2$, craters are not observed. However, the height of the bumps continues to increase. Peaks are observed to erupt in the fluence range 3.53 - 4.24 J/cm$^2$ and are observed with a maximum height of 182.50 nm with a substantial standard deviation as indicated by the error bars on the graph. The height of the rims in the fluence range 4.59 - 7.07 J/cm$^2$ does not follow a particular trend but fluctuates by about 30 nm. However, the diameter of the rim steadily increases with fluence, approaching the beam spotsize. Craters/moats reappear in the fluence region 4.94 - 7.07 J/cm$^2$, which increase in depth with laser pulse fluence.

FESEM of single laser pulse modified topologies at fluences 2.6 J/cm$^2$ and 2.8 J/cm$^2$, are shown in *fig.5(a)* and *(b)* respectively. In this fluence region FESEM reveals a bump of 1µm - 2µm in diameter with a hole (crater) in the laser affected region. The location of the hole within the processed region is random for each processed site and even absent in a few modified regions. Close examination of the FESEM micrographs (*fig.5(a)&(b)*), reveals a few very small openings in the processed region, along with the presence of clusters of expelled material which show *in bright contrast*. The FESEM micrographs, in *fig.5(a), (b)&(c)*, show a transition from a *low bump* (with indents and mounding at an even smaller size scale that has been referred to as ripples in prior laser processing studies), at the lower fluence range, to a *higher bump with a central jet*, above the single laser pulse threshold for bump formation. In *fig.5(c)*, an additional bump almost central to the bump is observed, which also has an opening, indicating a bubble bursting in an elastic material. At a fluence of 4.6 J/cm$^2$, a rim is observed to surround the burst bump topology, *as shown in fig. 5(d)*. The height of the bumps and rims along with the depth and diameter of the craters continues to increase with laser pulse fluence, as shown in fig. 5(e) & 5(f), for fluences 5.65 J/cm$^2$ and 7.07 J/cm$^2$, respectively. However, the frequency of observing an opening at the top of the bump sites reduces substantially at higher fluence of order 7.07 J/cm$^2$. It should be noted that quantitative height measurements are not possible by FESEM.

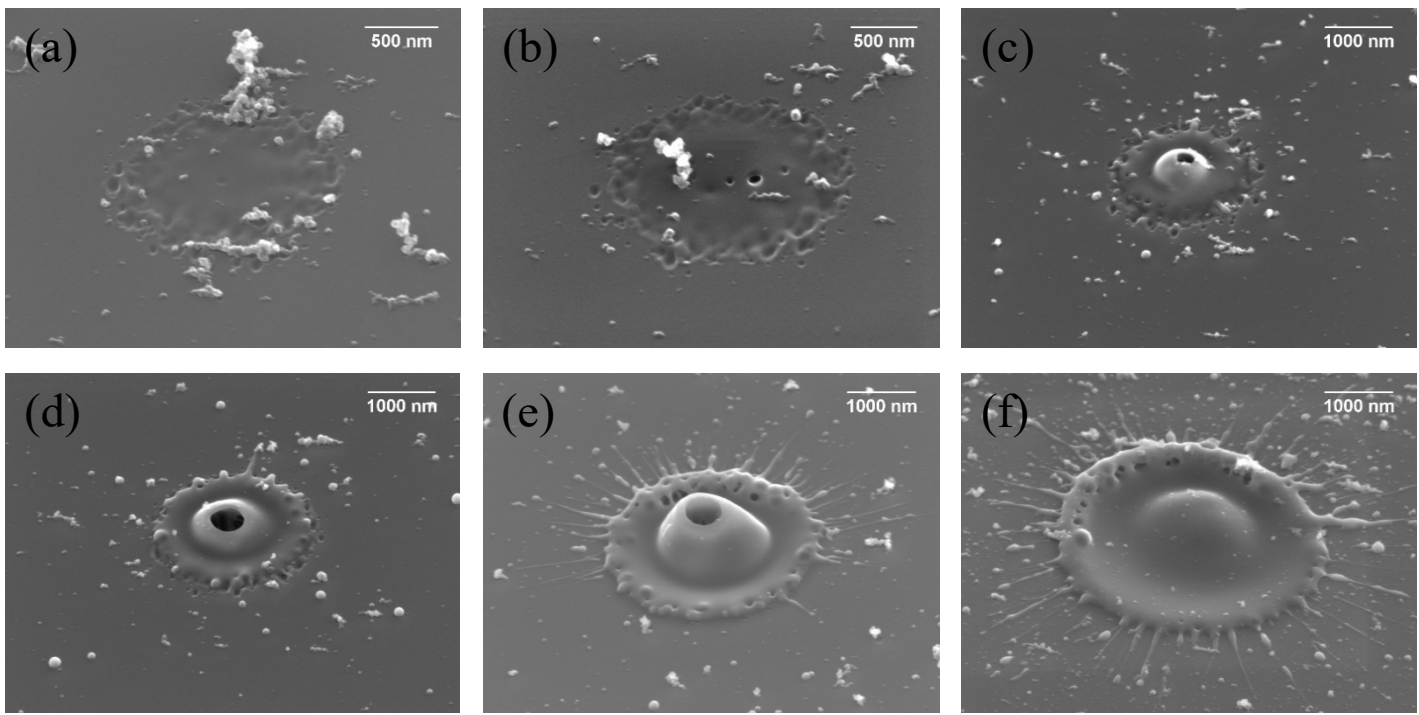

**Figure 5:** Field emission scanning electron microscope (FESEM) images at variable fluences in the range 2.4 J/cm$^2$-7.07 J/cm$^2$. Surface modification at fluence, (a) 2.65 J/cm$^2$, (b) 2.8 J/cm$^2$, (c) 3.88 J/cm$^2$, (d) 4.6 J/cm$^2$, (e) 5.65 J/cm$^2$, (f) 7.07 J/cm$^2$. Scale is shown in white bar and text with each image. Scale in (a), (b) is 500nm compared to 1000nm in (c), (d), (e), (f) because the size of the modified region in that fluence region is fairly small compared to higher fluence region.

Contrasting the two characterization techniques of OSP and FESEM confirms the unconventional progression of topologies with increasing single laser pulse fluence. Three dimensional maps of the modified topologies, obtained from OSP, depict the comprehensive analysis of the geometrical parameters, along with measured standard deviations from measuring 25 sites. Although FESEM allows high flexibility in magnification and gives higher lateral resolution for spatial features of the topologies, it is unable to provide a quantified three-dimensional map of the topologies. In contrast, OSP has the ability to scan large surface areas with sub nanometre vertical resolution (important in characterizing shallow craters and bumps) and is only constrained by the limited lateral resolution which does not resolve small openings and small indents/rippling in the laser processed topologies. FESEM imaging also requires sample coating for a dielectric, to manage charging effects, and is therefore a destructive characterisation technique. Whereas, OSP does not require any sample preparation and does therefore not alter the sample in any way.

**Discussion**

When a 300 µm muscovite sheet is irradiated with a single 150 fs duration laser pulse at 800 nm, focussed on the surface to a spotsize of ~6 µm, significant energy coupling occurs resulting in a systematic series of laser processed topologies, as the pulse fluence is increased. Volumetric removal of material (0.54 - 4.93 µm$^3$), is also observed with an increase in fluence, ~(5 – 7) J/cm$^2$. At 800 nm wavelength with a 150 fs laser pulse the dominant process of energy coupling in muscovite is expected to be non-linear multi-photon ionization of the crystal[14,15], i.e. six-photon ionization. However, we observe the low threshold for the surface modification by a single fs laser pulse, identified to be 2.4 J/cm$^2$. Conventionally, the laser fluence threshold for the surface modification is defined at the lowest fluence of observed visual modification at the surface, after which any further increase leads to ablated sites which deepen and/or widen with increasing fluence. However, in the case of muscovite, the mineral water [40–42] alters the modification mechanism upon impact of a fs pulse as compared to conventional dielectrics. As a consequence, the modified sites do not follow the traditional ablation progression upon increase in the fluence.

The smaller diameter of the modified sites near the *crater* threshold, in comparison to the 6 µm spotsize of the pulse, as shown in *fig.4b*, indicate that only the centre of the incident pulse modifies the small volume of 0.046 µm$^3$. We estimate the affected volume to encapsulate 2.4∗10$^7$ H$_2$O molecules that only require a small fraction of the incident pulse energy i.e. in nJ regime, to cause vaporization (Supplementary section E). From the small openings in the modified region as observed in the FESEM micrographs, *in fig. 5(a)&(b)*, we propose the cavitation of mineral water[43], leading to the formation of water vapour pockets[44,43] which both transfer thermal energy to the surrounding material, rendering it deformable, and expand to bursting (microexplosions[21]), escaping the processed region and giving rise to the openings. Clusters of debris, shown in bright contrast are proposed as some combination of the non-volatile component of mineral water[35] and small quantities of the elemental composition of the muscovite, or a subset thereof. The observation of smaller, unexploded bubbles around the central *popped* bump, supports the proposal of a mineral water assisted micro-bubbling[45,43] in the processed region. At higher fluence, *fig. 5(c),* the modified region depicts bubbling and formation of nano pores within the bump (Supplementary section B and F), similar to the modifications in ultrafast pulse processed polymers[46]. However, there are no signs of dissociation as in polymers. At this fluence muscovite modifies moderately like a polymer. The openings in the bump reveal fibre and nano-pore like arrangements inside the bumps, similar to laser processed polymers[46], accompanied by the signatures of molten material resetting. Expanding gas pockets, proposed to be primarily vaporised interlayer mineral water, drive the micro explosion of a hot muscovite jet perpendicular to the surface. Fast cooling thereafter freezes the material in a jet like structure[47]. Hence, the height of the jet shows a large range of values depending on where in its projection and collapse its re-solidification is captured (fig.4(a), 3.53-4.24 J/cm$^2$).

The higher fluence, *shown in fig.5(d), (e) & (f)*, leads to a softening of a larger volume of material in the processed region. The appearance of a rim of hot displaced material indicates, a further rise in lattice temperature[1]. It occurs along the boundary of the temperature differential[48] between the high temperature soft central processed region, and the cold surrounding material. Such a rim will shield the hot material flowing under the influence of surface bound shock waves[49]. At fluences larger than 3.88 J/cm$^2$ deep craters surrounded by a rim are common features in the modified topologies. As craters, they follow conventional expectation for fs laser ablation in dielectrics[1,50]. Material has been removed from the ablation site. However, bumps are still present inside these craters. The height of the bumps, as shown in *fig. 3(d)*, can reach ~130nm (*fig.4a*), with an opening at the top. Nanopores inside the bumps are still observed through the opening, hence, polymer like modification still occurs at such fluence. The increase in height of the bumps can be attributed, to the rise in lattice temperature, as evidenced by rim formation[1]. At highest fluence used in the study, i.e. 7.07 J/cm$^2$, shown in *fig.3(e) & 5(f)*, the topology follows the trend of a rim surrounding the crater with a bump inside it. The number of encapsulated $H_2O$ molecules estimated to be in the processed volume increases to $9.65*10^{10}$, however, the opening in the bumps are absent in all observations. It indicates the gas pockets are not able to escape the region. It is likely the dynamic re-solidification of hot material (following ablation), suppresses the bursting (effect of micro-explosion of gas pockets) in the larger affected volume.

## Conclusion

This is the first study to highlight a distinct sequence of topologies obtained by single fs laser pulse processing of the dielectric muscovite, a mica. Usually there is a state-of-the-art set of expectations for the systematics of the topology of laser processing of a dielectric using ultrafast (fs) pulses, as the fluence of the pulses increases. This is informed by the physics of ultrafast light matter interactions. Significant evidence of modification is obtained with just a single 150 fs, 800 nm, 6 micron spotsize laser pulse at a fluence 2.4 J/cm$^2$ or greater. Usually multiple laser pulses are used because current knowledge suggests they are required. The systematic variation in topology of the single laser pulse processed sites indicates several physical mechanisms may be sequentially contributing to the results. With an increase in pulse fluence the modification mechanism in muscovite may be transitioning from escaping gas pockets to polymer like cavitation and bubbling, followed by micro explosion, and then conventional ablation accompanied by cavitation, bubbling and micro explosion.

The unconventional response of muscovite sheet to an ultrafast pulse, is attributed to its atypical layered mineral structure incorporating interlayer mineral water. It is proposed the interlayer mineral water changes to vapour after absorbing pulse energy and the vapour transfers energy to the lattice. Nucleation of $H_2O$ gas bubbles followed by volatile burst(s) in the processed region, result in the formation of bumps above the surface. Hence, we propose a first instance of $H_2O$ assisted laser modification in a transparent mineral by single fs pulse irradiation.

The system is a challenging one to model theoretically and these experimental results will hopefully motivate experts in molecular dynamics to start addressing a layered crystalline muscovite, or an appropriate simpler crystal in the first instance, with interlayer water. Such a material structure emerges as one that might be fabricated and explored for its science interest and potential usefulness.

## Method and Instruments

The laser system used in this study was a regeneratively amplified ultrafast Ti: Sapphire laser (Coherent RegA 9000), emitting pulses at a wavelength of 800 nm with a duration of 150 fs, a maximum repetition rate of 100 kHz and a maximum average output power of 0.212 W. The pulses were focused using a 10× magnification objective (Olympus ULWD), resulting in a 1/e2 spot size of around 6 μm in diameter. In our experiments, the repetition rate was lowered to 10 Hz and the Position Synchronised Output (PSO) of the positioning stages (see below) was used to trigger the laser to ensure single pulse interaction at the pre-defined locations on the surface of the sample. The pulse energy incident on the sample was adjusted by rotating a half-wave plate in front of a polariser and was monitored using a powermeter (Thorlabs PM100A). The polarization of the laser pulses incident on the sample was changed to circular by a quarter-wave plate to exclude polarization-dependent effects.

The sample was positioned by a computer-controlled precision XYZ stage (Aerotech ABL1500s). Alignment was aided by a vision system incorporated into the focusing arm of the laser path, which enabled real-time monitoring. A freshly cleaved muscovite sheet of dimension 25 mm $x$ 25 mm $x$ 300 microns from Axim Mica, Robbinsville, NJ, USA was used for the laser processing.

Suraface marking of the laser processed pattern to be characterized was performed in a 10 × 100 matrix pattern with ten rows corresponding to different single pulse fluence levels from 1.76 J/cm$^2$ – 7.07 J/cm$^2$. The feed rate of the translation stage is set (with reference to the 10 Hz repetition rate) to equally space individual spots, being exposed to a single laser pulse to allow statistical analysis of the resulting interaction. Different modification regimes were investigated by varying the laser power, starting with the highest achievable laser pulse fluence, well-above the surface modification threshold, and then reducing it sequentially until it was below the modification threshold.

The laser-induced modifications were characterized using a Bruker-AXS NT-9800 optical surface profiler (OSP) (working principle is described in supplementary information **section F**) and a JEOL JSM 7100F – Field Emission Scanning Electron Microscope (FESEM).

For OSP characterization, a 50× and a 115× magnification objective was used in combination with a 1.0× field of view lens. With the 50× objective the field of view is 140 μm x 100 μm, and with 115× it is 55 μm x 40 μm. The samples (laser processed muscovite sheets) were mounted on a manual stage with an X/Y translation range of ± 50.8mm (± 2in.) and ± 4° tip/tilt. OSP characterization is performed without any post processing of the sample after laser processing.

The FESEM micrographs were obtained at an operating voltage of 15 kV. FESEM micrographs were recorded at a magnification of 33000, 30000, 27000 and 18000 times and at a working distance of 24.00 mm, 23.9 mm, 23.9 mm and 24.4 mm, respectively. A turbomolecular pumped coater (a vacuum evaporative coating unit, Quorum Q 150T) was used to coat the muscovite laser processed sample with a thin carbon layer prior to FESEM imaging.

## Acknowledgement

We thank Dr. Chris Marjo and Dr. Bill Gong, (Solid State & Elemental Analysis Unit, University of New South Wales (UNSW), Sydney, Australia) for assisting with X-Ray fluorescence (XRF) characterization. We thank Dr. Benjamin Johnston, (OptoFab (ANFF) Macquarie University, Sydney, Australia) for expertise in muscovite laser processing, and Dr. Chao Shen, (Microscopy unit, Faculty of Science, Macquarie University, Sydney, Australia) for assistance with field emission scanning electron microscopy (FESEM) characterization.

## Appendices A–F. Supplementary data

Supplementary data to this article is attached.

**Single femtosecond laser pulse interaction with mica**

*Saurabh Awasthi[1*], Douglas J. Little[1], A. Fuerbach[1], D.M. Kane[1*]*

1. MQ Photonics research centre,

Department of Physics and Astronomy, Macquarie University,

Sydney, NSW-2109

[*]Corresponding authors E-mail : - deb.kane@mq.edu.au ; saurabh.awasthi@hdr.mq.edu.au

# Supplementary Information

The information supplementing the main text of the manuscript entitled '*Single femtosecond laser pulse interaction with mica*' is divided into 6 sections (A-F). **Section A** contains the data of depth and diameter for the laser processed sites at different fluences as obtained from the optical surface profiler (OSP). **Section B** shows the field emission electron microscope (FESEM) images of the characterized sites at the same fluence levels. **Section C** contains the data from the X-ray fluorescence measurements that were done to identify the $H_2O$ content in the muscovite sheets used in the study. **Section D** shows the calculation of the number of the $H_2O$ molecules that are most likely encapsulated in the laser pulse affected volume. **Section E** introduces that modification observed in the muscovite using FESEM images is indicative of polymer-like laser processing for a particular range of fluences. Finally, **Section F** presents the schematic and working principle for the optical surface profiler (OSP) used in this study.

# Section-A
# (Optical Surface Profiler Characterization)

This section contains plots and tables showcasing the geometrical parameters of average height and average diameter of the laser processed modified sites for each pulse fluence, along with their standard deviations. For the sake of clarity, the geometrical data for each individual identifier (crater, bump, rim and jet) is presented individually. Figures 6 - 9 also include the source data that was used to create figure 4 in the main text of the paper.

| Incident pulse Fluence | Crater (mean) (±S.D.) | |
|---|---|---|
| | Depth (nm) | Diameter (micron) |
| **2.47 J/cm²** | **3.84±1.07** | **1.37±0.16** |
| **2.65 J/cm²** | **3.38±1.04** | **1.51±0.18** |
| **2.82 J/cm²** | **10.04±1.66** | **1.31±0.09** |
| **3.00 J/cm²** | **11.83±3.62** | **1.73±0.24** |
| **3.18 J/cm²** | - | - |
| **3.53 J/cm²** | - | - |
| **3.88 J/cm²** | - | - |
| **4.24 J/cm²** | - | - |
| **4.59 J/cm²** | **141.82±30.07** | - |
| **4.94 J/cm²** | **16.72±10.82** | - |
| **5.30 J/cm²** | **30.10±10.23** | - |
| **5.65 J/cm²** | **76.00±24.69** | - |
| **6.01 J/cm²** | **90.90±21.47** | - |
| **6.36 J/cm²** | **126.61±14.82** | - |
| **7.07 J/cm²** | **138.72±18.14** | - |

(a)

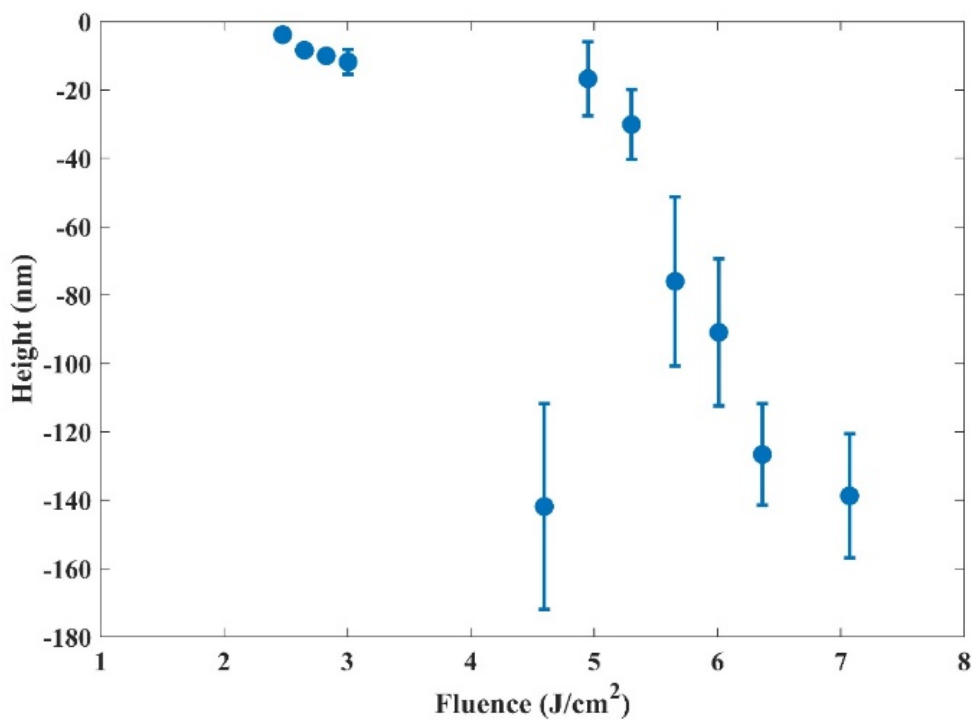

(b)

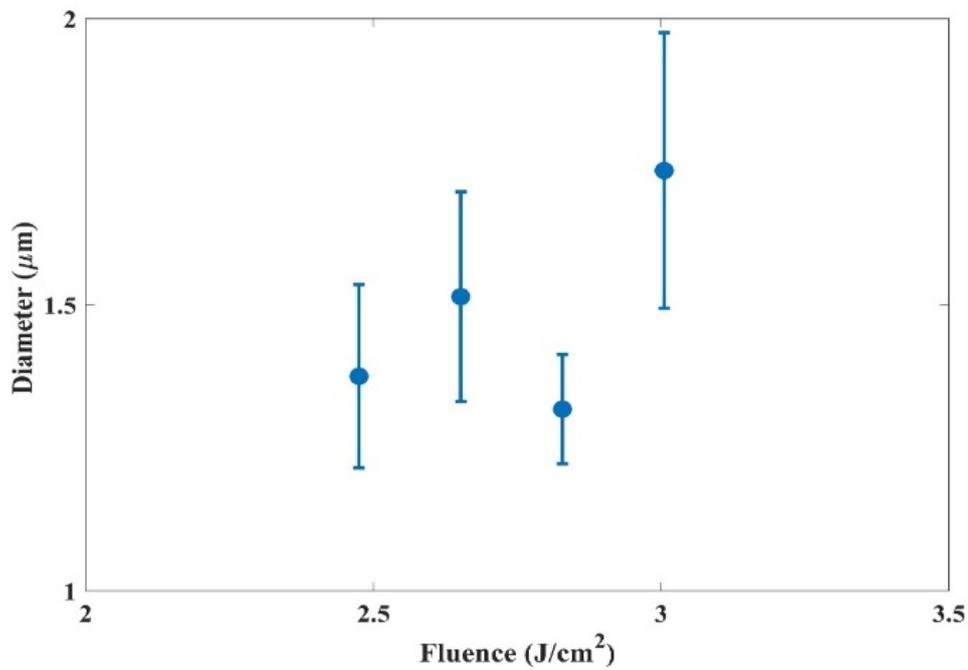

(c)

**Figure 6:** Geometrical parameters of crater type modifications at various fluences as obtained by the optical surface profiler (OSP). A negative value in height indicates that the modification is beneath the basal surface plane. The depth of the craters is observed to monotonically increase with fluence, except for an anomaly at 4.59 J/cm². No modification is observed below the surface in the fluence range 3.18-4.24 J/cm². The diameter of the craters in the fluence range > 4.59 J/cm² is not considered here as the modified craters are encircled by rims in this case and the diameter of encircling rim quantifies the diameter of the overall modification.

| Incident pulse Fluence | Bump (mean) (± S.D.) | |
|---|---|---|
| | Height (nm) | Diameter (micron) |
| **2.47 J/cm²** | - | - |
| **2.65 J/cm²** | - | - |
| **2.82 J/cm²** | 6.75±0.89 | 7.85±0.56 |
| **3.00 J/cm²** | 11.08±2.74 | 8.68±0.82 |
| **3.18 J/cm²** | 26.60±2.06 | 13.02±1.42 |
| **3.53 J/cm²** | 53.28±14.67 | 12.06±0.84 |
| **3.88 J/cm²** | - | 3.88±0.36 |
| **4.24 J/cm²** | - | 3.99±0.45 |
| **4.59 J/cm²** | 137.44±15.40 | 14.15±0.51 |
| **4.94 J/cm²** | 39.14±10.21 | - |
| **5.30 J/cm²** | 47.89±24.59 | - |
| **5.65 J/cm²** | 67.15±15.74 | - |
| **6.01 J/cm²** | - | - |
| **6.36 J/cm²** | - | - |
| **7.07 J/cm²** | - | - |

(a)

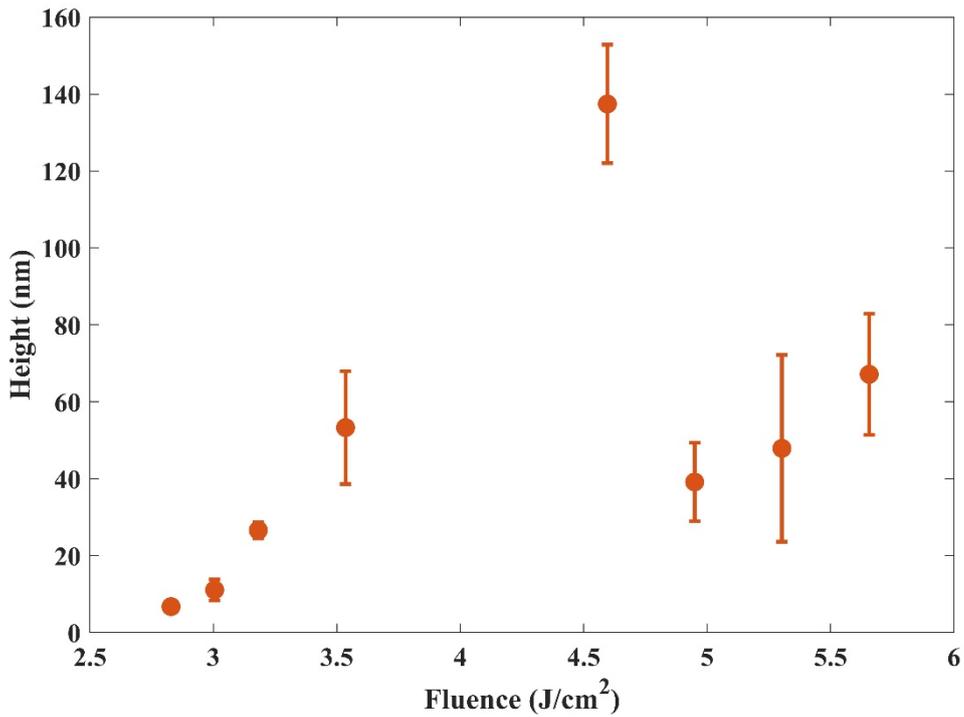

(b)

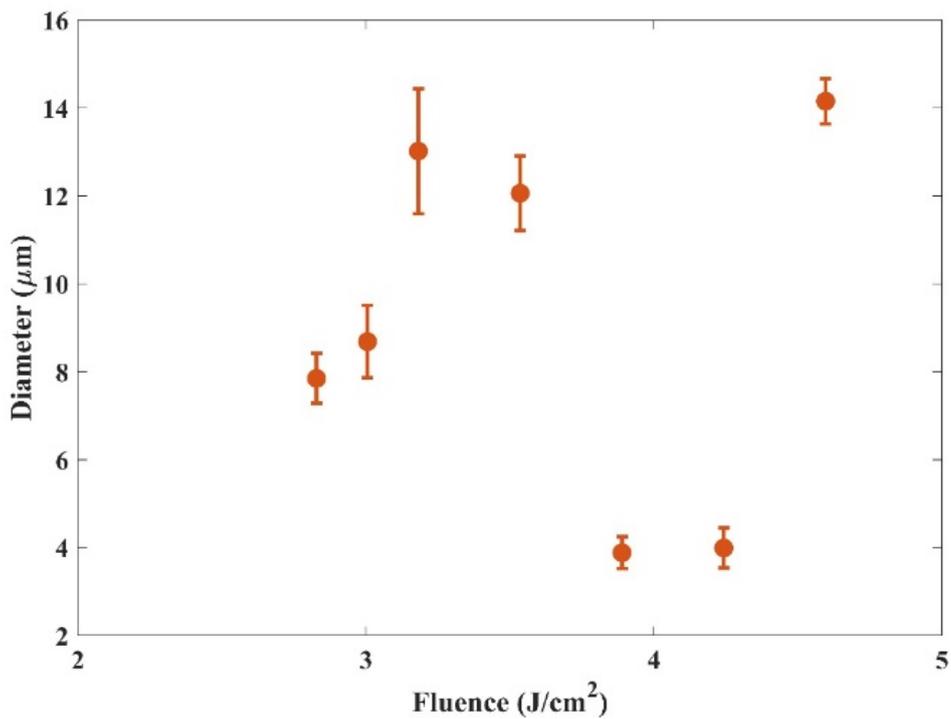

(c)

**Figure 7:** Geometrical parameters of bump-type modifications at various fluences, as obtained by the optical surface profiler (OSP). The height of the bumps increases monotonically in the fluence range 2.82-4.59 J/cm$^2$, then drops at a fluence of 4.94 J/cm$^2$, and again increases monotonically thereafter. The diameters of the bumps follow a similar, yet less clear trend. Diameters in the fluence region > 4.94 J/cm$^2$ are not considered due to the appearance of rims.

| Incident pulse Fluence | Rim (mean) (±S.D.) | |
|---|---|---|
| | Height (nm) | Diameter (micron) |
| **2.47 J/cm²** | - | - |
| **2.65 J/cm²** | - | - |
| **2.82 J/cm²** | - | - |
| **3.00 J/cm²** | - | - |
| **3.18 J/cm²** | - | - |
| **3.53 J/cm²** | - | - |
| **3.88 J/cm²** | **39.23±5.84** | - |
| **4.24 J/cm²** | **50.43±3.95** | - |
| **4.59 J/cm²** | **58.29±12.43** | **2.49±0.05** |
| **4.94 J/cm²** | **24.06±4.36** | **2.68±0.28** |
| **5.30 J/cm²** | **28.87±6.46** | **2.80±0.04** |
| **5.65 J/cm²** | **53.00±8.69** | **2.83±0.10** |
| **6.01 J/cm²** | **45.56±11.99** | **3.14±0.07** |
| **6.36 J/cm²** | **66.92±6.83** | **3.33±0.11** |
| **7.07 J/cm²** | **31.37±19.87** | **4.52±0.26** |

(a)

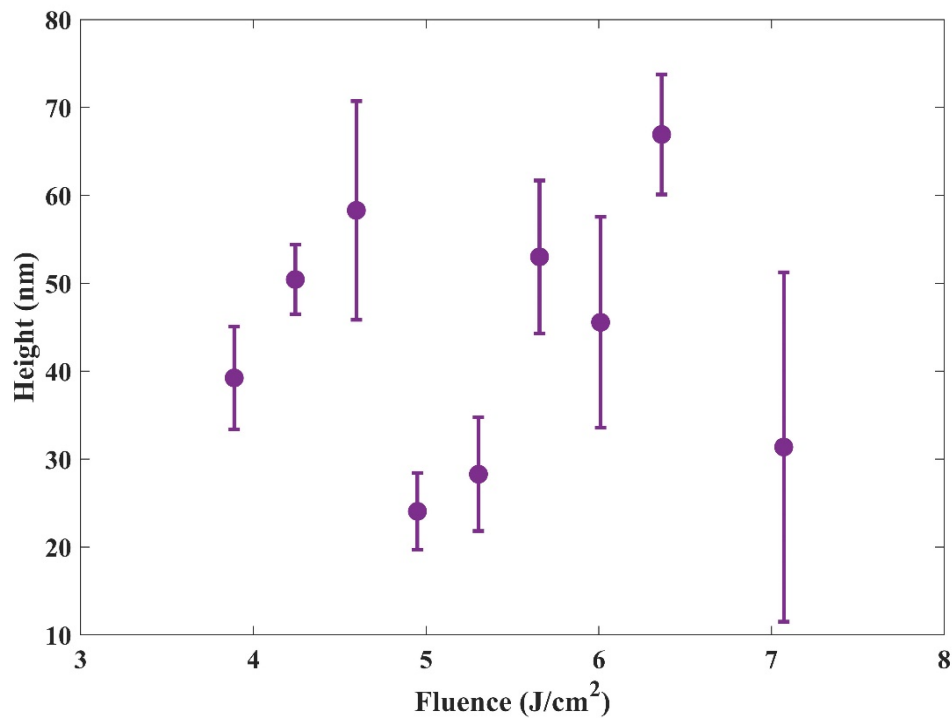

(b)

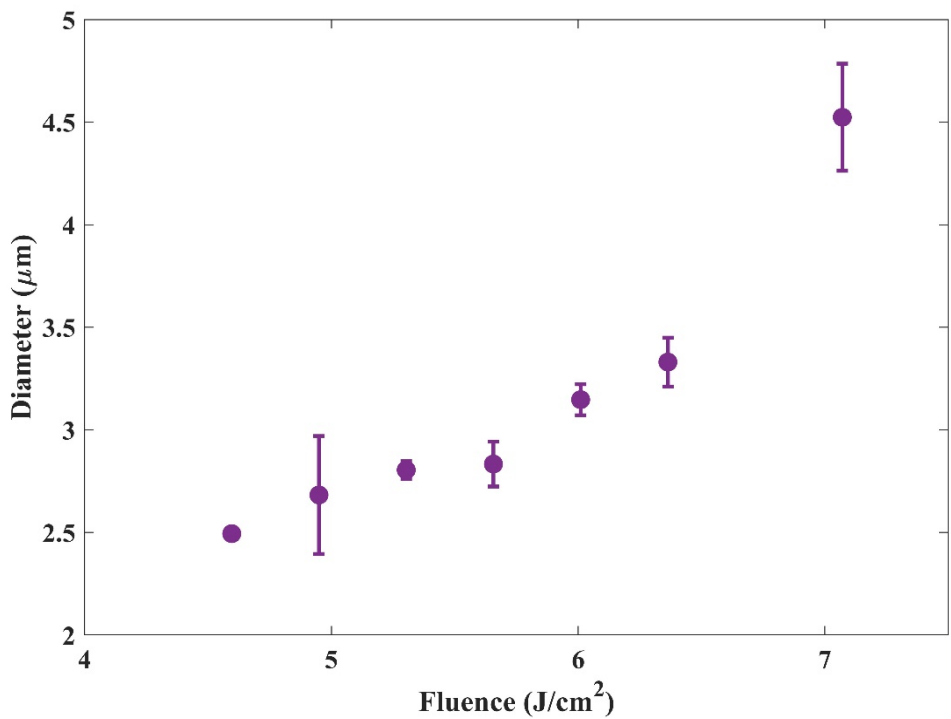

(c)

**Figure 8:** Geometrical parameters of rim type modifications at various fluences as obtained by the optical surface profiler (OSP). Rims start to appear at a fluence of 3.88 J/cm$^2$, surrounding the sharp jet and encircling the modified surface throughout the fluence region of interest. The height of the rims does not follow a particular trend as a function of fluence, but remains in the range 24.06 - 66.92 nm. The diameter of the rims increases monotonically with fluence and approaches the diameter of the incident laser beam.

| Incident pulse Fluence | Jet (mean) (± S.D.) |
|---|---|
|  | Height (nm) |
| 2.47 J/cm$^2$ | - |
| 2.65 J/cm$^2$ | - |
| 2.82 J/cm$^2$ | - |
| 3.00 J/cm$^2$ | - |
| 3.18 J/cm$^2$ | - |
| 3.53 J/cm$^2$ | 140.67±43.90 |
| 3.88 J/cm$^2$ | 182.50±113.73 |
| 4.24 J/cm$^2$ | 174.92±76.23 |
| 4.59 J/cm$^2$ | - |
| 4.94 J/cm$^2$ | - |
| 5.30 J/cm$^2$ | - |
| 5.65 J/cm$^2$ | - |
| 6.01 J/cm$^2$ | - |
| 6.36 J/cm$^2$ | - |
| 7.07 J/cm$^2$ | - |

(a)

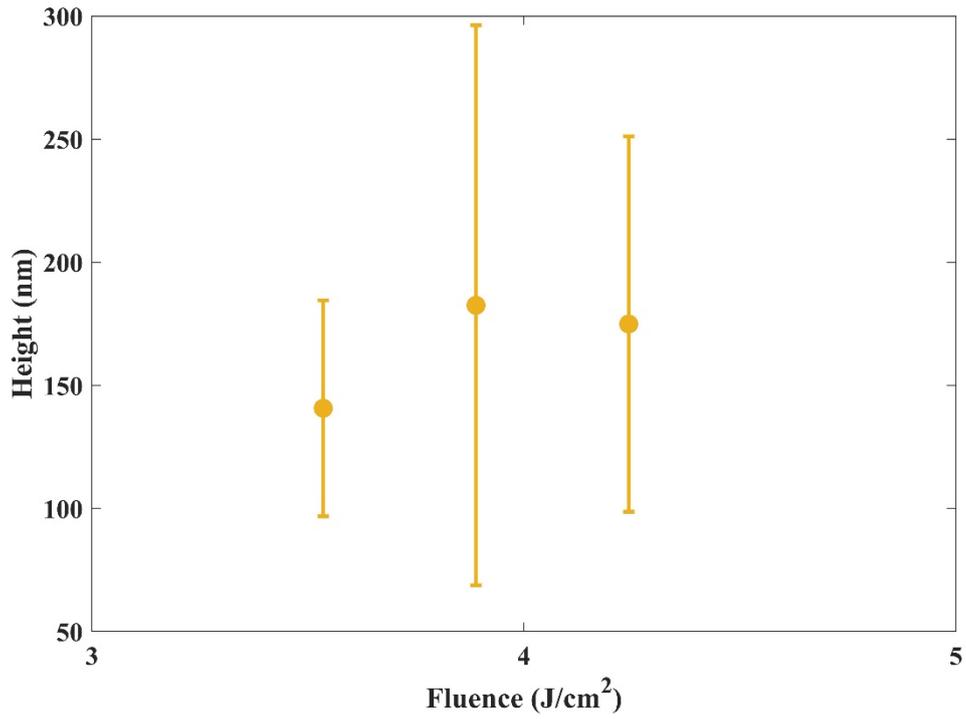

(b)

**Figure 9:** Height of the unique jet like modifications at the surface. Jets can be visualized as sharp and steep protrusions at the laser processed basal surface layer, hence no diameter values are given. Jets are observed to appear in the fluence range 3.53 - 4.24 J/cm$^2$ and are observed to reach a maximum height of approx. 300 nm.

# Section-B

# (Field Emission Scanning Electron Microscope Characterization)

This section contains high magnification FESEM images of the laser processed muscovite sites that supplement figure 5 in the main text and highlight micro-bubbling, bursting and spattering as discussed in the main text.

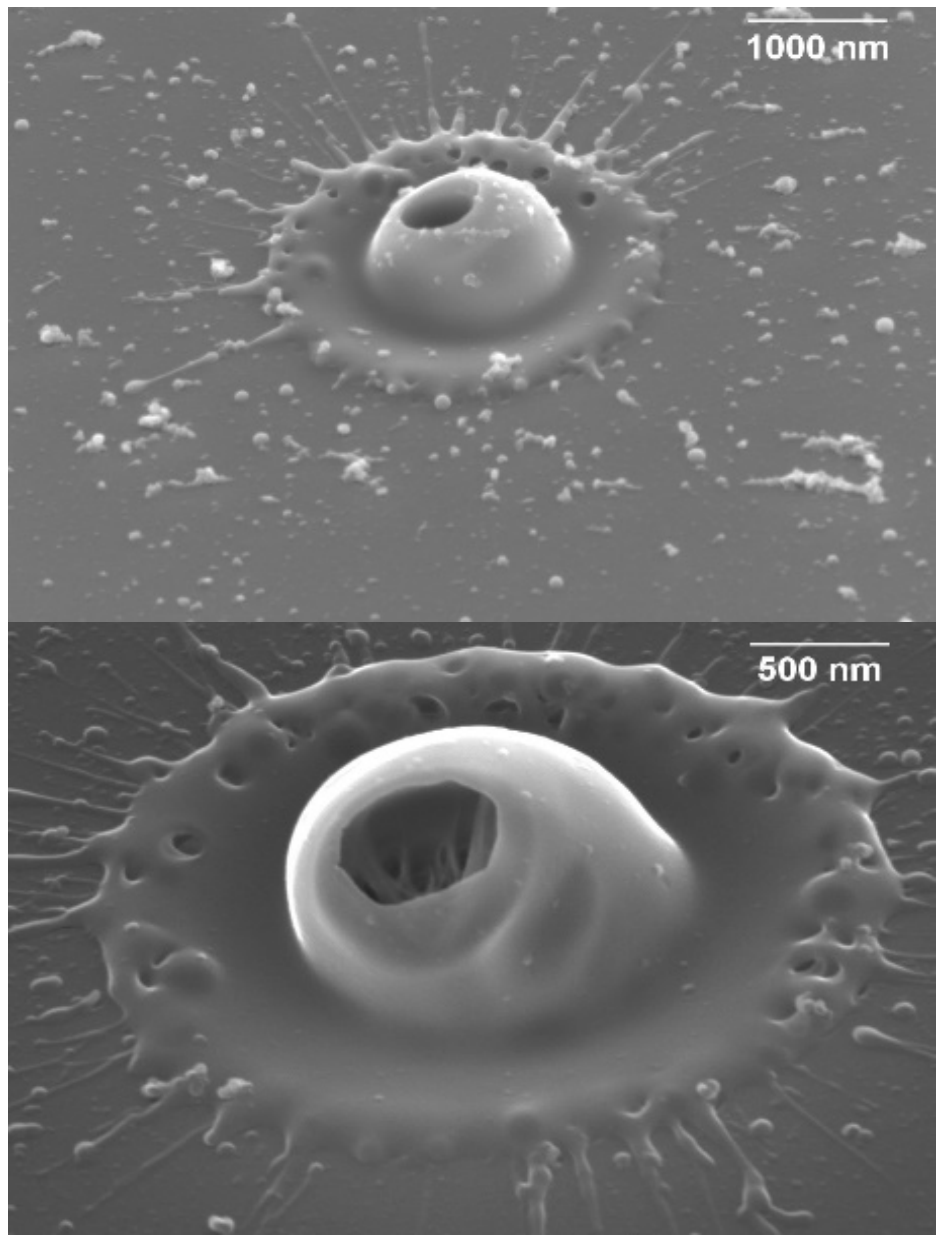

**Figure 10:** FESEM of the laser modified surface region at fluence 4.59 J/cm$^2$. (a) is imaged at a magnification of 18,000 x while the magnification is increased to 35,000 x in (b). In (a), the modified region is observed to have a central bump topology with an opening at its top surrounded by a rim. The inside of the central bump topology can be seen in (b) which confirms the hollow nature of the topology with a scaffold like arrangement. The higher magnification image in (b) also accentuates the micro-bubbling along the rim surrounding the central topology. Signatures of spatter along the rim coming from burst micro bubbles support the mechanism of micro explosion assisted by H$_2$O. Close examination of (a) also reveals the presence of ejected surface debris.

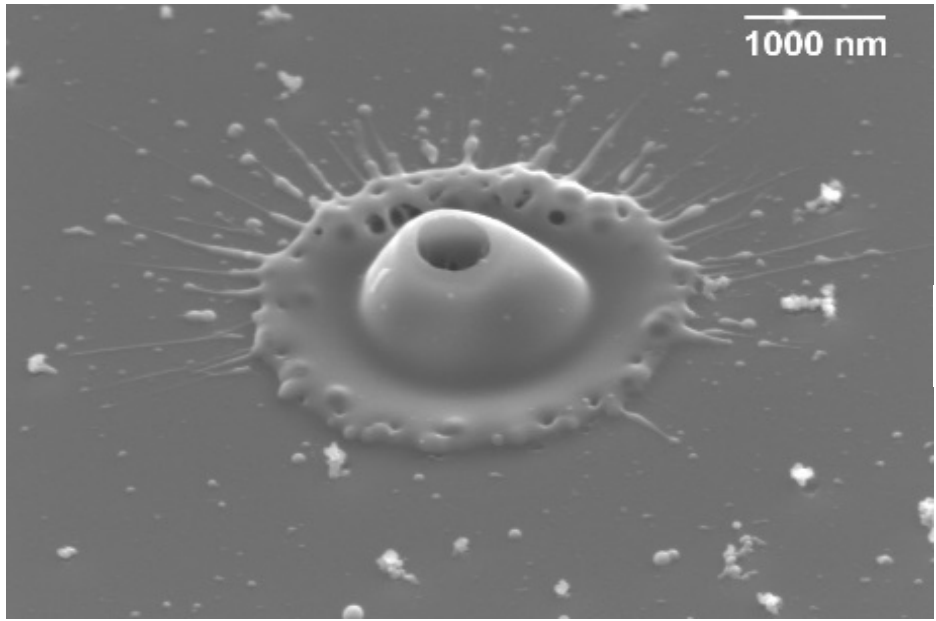

(a)

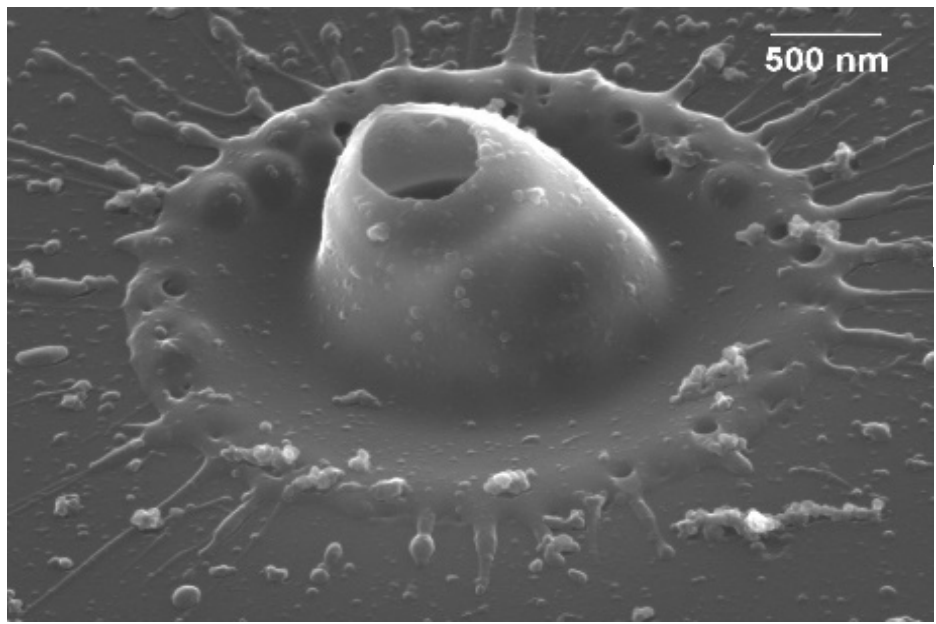

(b)

**Figure 11:** FESEM micrograph of the laser modified surface region at fluence 4.94 J/cm². (a) and (b) are imaged at magnifications 18,000 x and 30,000 x, respectively. (a) reveals a smooth central bump topology with an opening at its top surrounded by a rim with micro-bubbles. However, higher magnification and change in contrast in (b), leads to the observation of settled small debris particles at the outer surface of the bump. The higher magnification in (b) also highlights the cluster of debris in the vicinity of the processed region. We suspect the debris cluster to be ejected from the processed region to be a compositional subset of muscovite mica.

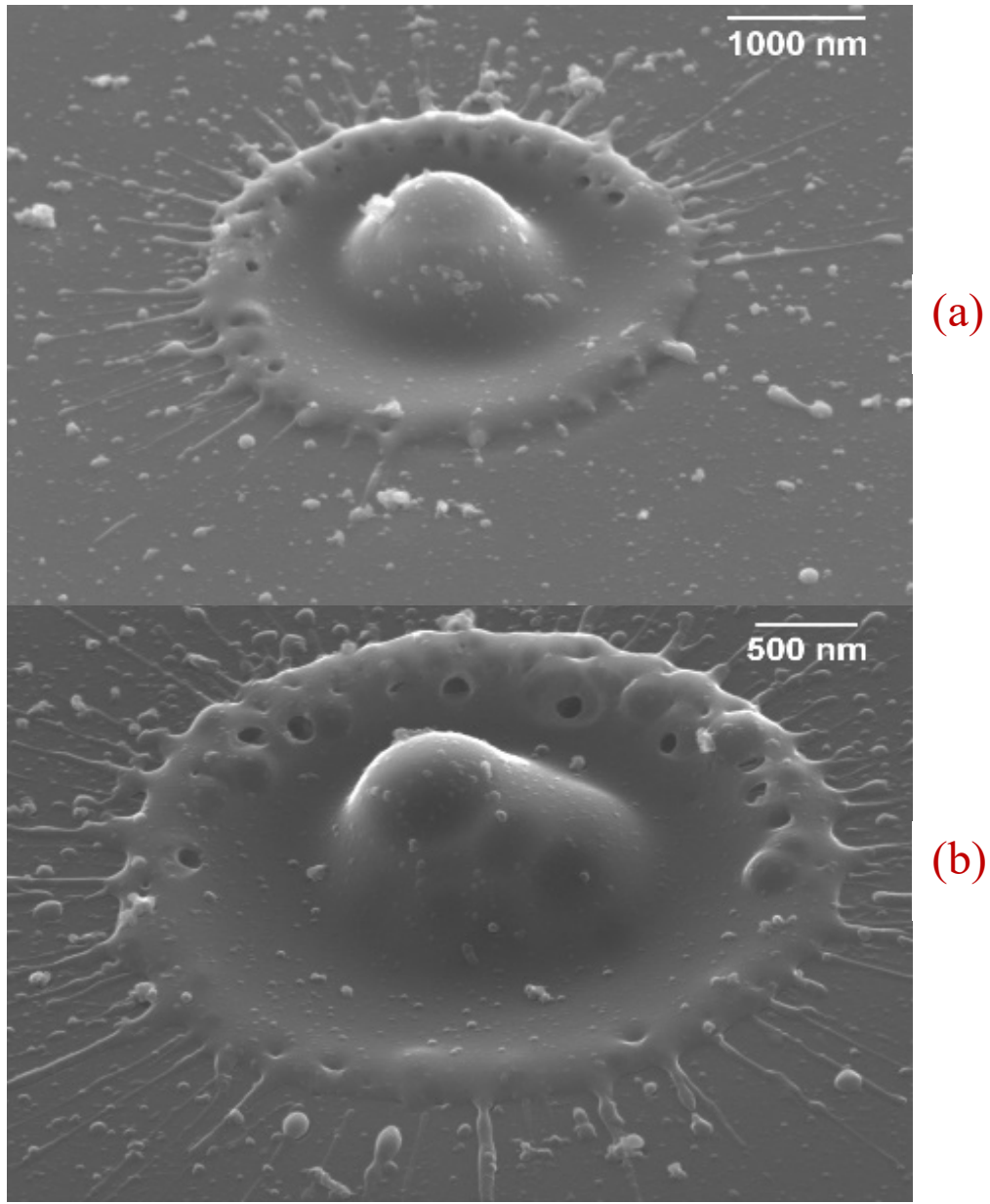

**Figure 12:** FESEM images of the processed site at fluence 6.01 J/cm$^2$, reveal the rim to surround the central bump topography. However, at this fluence range most of the bumps do not have openings at the top. Signature of spatter and micro-bubbling along the rim remains visible.. (a) and (b) are imaged at magnifications 18,000 x and 27,000 x, respectively.

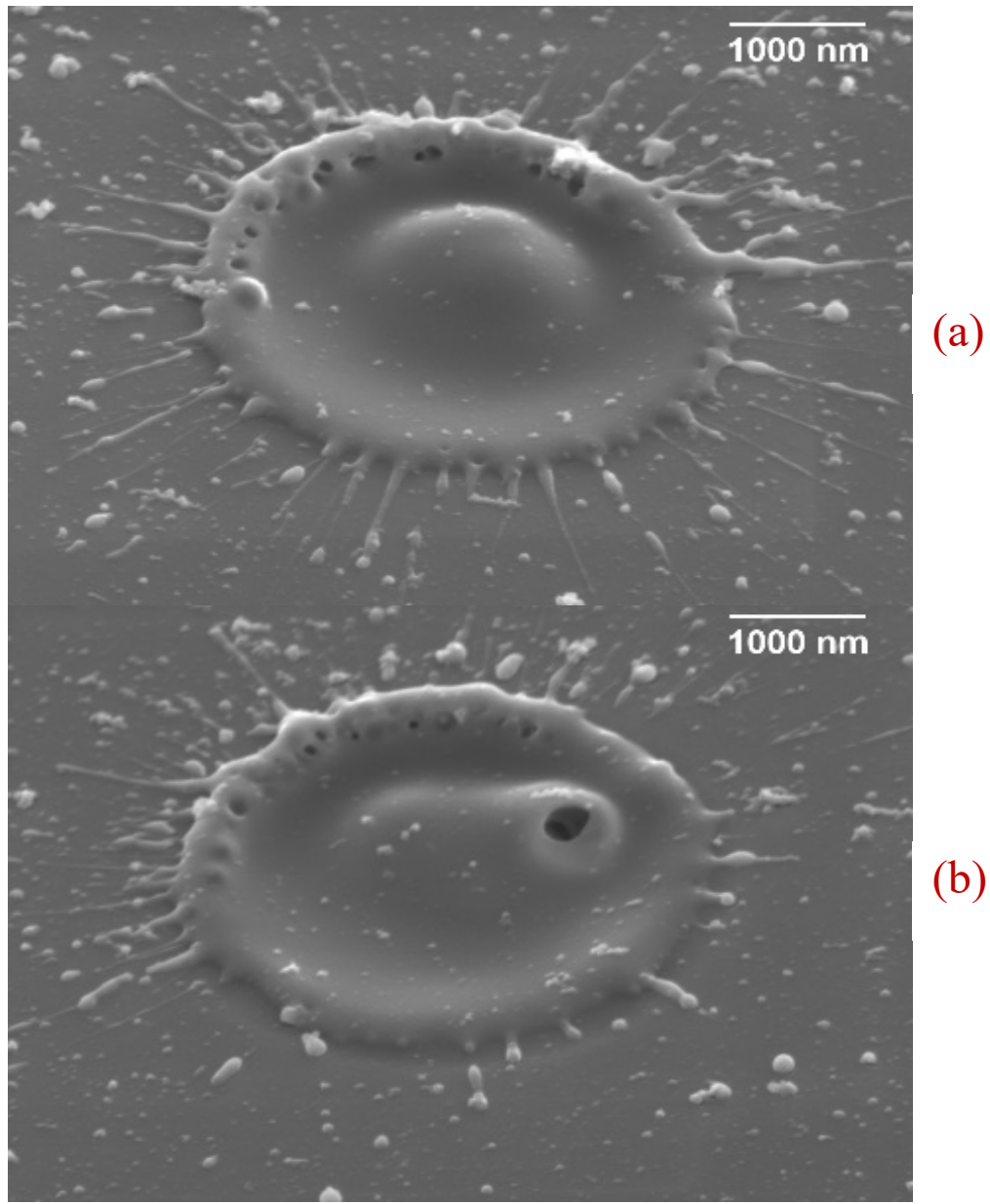

**Figure 13:** The FESEM images in (a) and (b) show the popped and un-popped central configuration at the same fluence of 7.01 J/cm$^2$. The reason for such anomaly is attributed to the fact that mineral muscovite has non-uniform structural composition along the processed surface, resulting in variable modifications at some processed sites. The jet like conical configuration in (b) has the opening discernibly off centre of the processed region. This could be the result of collapse of the modified jet in a particular direction prior to the re-solidification.

# Section-C
# (X-Ray Fluorescence Characterization)

This section contains the elemental analysis of the 300μm thick muscovite sheets weighting 3.68 grams each used in the study as obtained from the x-ray fluorescence measurement. It identifies the weight percentage of the individual elements comprising the muscovite sheets. Data present in this section is the source for figure 2 in the main text.

| Oxide wt.% | Muscovite Sample 1 | Muscovite Sample 2 |
|---|---|---|
| $SiO_2$ | 45.89 | 45.37 |
| $TiO_2$ | 0.07 | 0.07 |
| $Al_2O_3$ | 36.73 | 36.86 |
| $Fe_2O_3$ | 1.28 | 1.30 |
| $Mn_3O_4$ | 0.01 | <0.01 |
| MgO | 0.31 | 0.33 |
| CaO | <0.01 | <0.01 |
| $Na_2O$ | 1.01 | 1.01 |
| $K_2O$ | 10.25 | 10.23 |
| $P_2O_5$ | 0.02 | 0.01 |
| $SO_3$ | <0.01 | <0.01 |
| $Cr_2O_3$ | <0.01 | <0.01 |
| $ZrO_2$ | <0.01 | <0.01 |
| SrO | <0.01 | <0.01 |
| CuO | <0.01 | <0.01 |
| ZnO | <0.01 | <0.01 |
| NiO | <0.01 | <0.01 |
| BaO | 0.19 | 0.17 |
| PbO | <0.01 | <0.01 |
| F | ND | ND |
| L.O.I. | 4.72 | 4.72 |
| **TOTAL** | 100.48 | 100.07 |

**Figure 14:** X-ray fluorescence (XRF) elemental analysis data for the 300μm muscovite sheets. Two identical sheets from the same batch were used in the analysis for statistical validity. The Weight of the sheet was 3.68 grams.

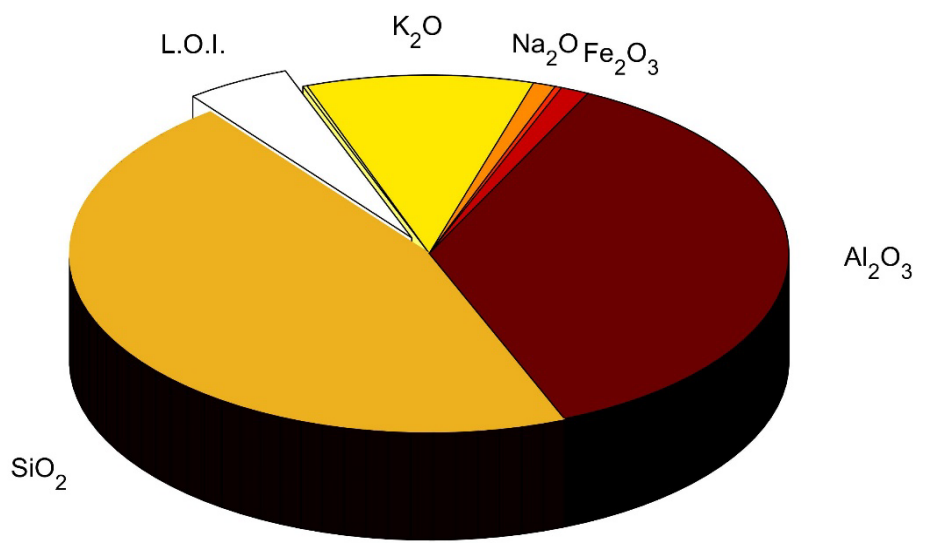

**Figure 15:** XRF elemental analysis data visualized as pie-plot. Si, Al and K are the major contributors to the elemental composition of the muscovite sheet in the study. The loss on ignition (L.O.I) component in the study is attributed to $H_2O$, as it is the most probable candidate to be lost on heating and the weight-% estimate is similar to previously reported data. Hence, $H_2O$ contributed 4.72 % of the total composition.

# Section-D

# Computation of encapsulated $H_2O$ molecules in the affected region and energy required to vaporize them.

This section shows the calculation to identify the number of water molecule that are most likely present in the laser pulse affected region in the muscovite. The calculation is built on the previous studies stating probabilistic models for possible number of molecules per unit cell and XRF data presented in the previous data. Data from this section is used to estimate the number of encapsulated water molecules in the processed region and the estimated laser energy required for its vaporization.

# Water molecule per unit cell calculation

Weight of muscovite sheet = 3.68 grams

Percentage of water in a sheet = 4.72 %

Dimension of the sheet = $51mm * 127mm * 300\mu m$

Water by weight in a sheet = $\frac{4.72}{100} * 3.68 = 0.173696 \; grams$

Number of individual layers in a sheet = $\frac{200*10^{-6}}{1.29*10^{-9}} = 155.0387596 * 10^3 \; layers$

Water by weight per layer = $\frac{0.173696}{155.0387596*10^3} = 1.12033920 * 10^{-6} \frac{grams}{sheet}$

Volume of a layer = $51 * 10^{-3} * 127 * 10^{-3} * 1.29 * 10^{-9} = 8355.33 * 10^{-15} \; m^3$

Volume of a unit cell at 20º C = $9.219 * 10^{-12} \; m^3$

Number of unit cells in a layer = $\frac{8355.33*10^{-15}}{9.219*10^{-28}} = 906.31630328 * 10^{13}$

Molecular mass of a water molecule = 18g/mol

1 gram of water contains = $\frac{6.022*10^{23}}{18} = 3.34455 * 10^{22} molecules$

Number of water molecules per layer = $1.12033920 * 10^{-6} * 3.3455 * 10^{22}$

$$= 3.74809473 * 10^{16}$$

Water molecule per unit cell = $\frac{3.74809473*10^{16}}{906.31630328*10^{13}} = 0.0041355526 * 10^3 \approx 4.135 \; molecules$

| Fluence | Depth of the topology | Diameter of the topology | Volume of crater | Number of layers affected | Number of unit cells | Number of water molecules | Molecules to grams | Enenrgy required |
|---|---|---|---|---|---|---|---|---|
| (J/cm2) | (µm) | (µm) | cubic µm |  |  |  | (g) | (nJ) |
| 2.47E+00 | 3.84E-03 | 1.37E+00 | 5.66E-03 | 2.98E+00 | 6.14E+06 | 2.45E+07 | 7.34E-16 | 1.66E-03 |
| 2.65E+00 | 3.38E-03 | 1.51E+00 | 6.05E-03 | 2.62E+00 | 6.56E+06 | 2.62E+07 | 7.85E-16 | 1.77E-03 |
| 2.82E+00 | 1.00E-02 | 1.31E+00 | 1.35E-02 | 7.78E+00 | 1.47E+07 | 5.87E+07 | 1.75E-15 | 3.97E-03 |
| 3.00E+00 | 1.18E-02 | 1.73E+00 | 2.78E-02 | 9.17E+00 | 3.01E+07 | 1.21E+08 | 3.61E-15 | 8.15E-03 |
| 3.18E+00 |  |  | 0.00E+00 | 0.00E+00 | 0.00E+00 | 0.00E+00 | 0.00E+00 | 0.00E+00 |
| 3.53E+00 |  |  | 0.00E+00 | 0.00E+00 | 0.00E+00 | 0.00E+00 | 0.00E+00 | 0.00E+00 |
| 3.88E+00 |  |  | 0.00E+00 | 0.00E+00 | 0.00E+00 | 0.00E+00 | 0.00E+00 | 0.00E+00 |
| 4.24E+00 |  |  | 0.00E+00 | 0.00E+00 | 0.00E+00 | 0.00E+00 | 0.00E+00 | 0.00E+00 |
| 4.59E+00 | 1.42E-01 | 2.49E+00 | 6.90E-01 | 1.10E+02 | 7.49E+08 | 2.99E+09 | 8.95E-14 | 2.02E-01 |
| 4.94E+00 | 4.67E-02 | 2.68E+00 | 2.63E-01 | 3.62E+01 | 2.86E+08 | 1.14E+09 | 3.42E-14 | 7.73E-02 |
| 5.30E+00 | 3.01E-02 | 2.80E+00 | 1.85E-01 | 2.33E+01 | 2.01E+08 | 8.04E+08 | 2.40E-14 | 5.43E-02 |
| 5.65E+00 | 7.60E-02 | 2.83E+00 | 4.78E-01 | 5.89E+01 | 5.18E+08 | 2.07E+09 | 6.20E-14 | 1.40E-01 |
| 6.01E+00 | 9.09E-02 | 3.14E+00 | 7.04E-01 | 7.05E+01 | 7.63E+08 | 3.05E+09 | 9.13E-14 | 2.06E-01 |
| 6.36E+00 | 1.27E-01 | 3.33E+00 | 1.10E+00 | 9.81E+01 | 1.20E+09 | 4.78E+09 | 1.43E-13 | 3.23E-01 |
| 7.07E+00 | 1.39E-01 | 4.52E+00 | 2.22E+00 | 1.08E+02 | 2.41E+09 | 9.65E+09 | 2.89E-13 | 6.53E-01 |

**Figure 16:** Above table approximates the number of water molecules that could be encapsulated in the laser pulse affected volume of the processed site and energy required to bubble it. Volume of the processed sites (Only fluences resulting in crater sites are considered in this calculation) is calculated by integrating the volume of individual pixel over the three dimensional surface as observed by the optical surface profiler. Number of affected layers are computed based on the depth of crater as observed by the OSP.

# Section-E
# (Polymer like modification in muscovite)

This section contains the FESEM micrographs depicting probable polymer like modification in the muscovite at certain fluences. These images are presented to supplement the possibility highlighted in the discussion and conclusion sections of the main manuscript.

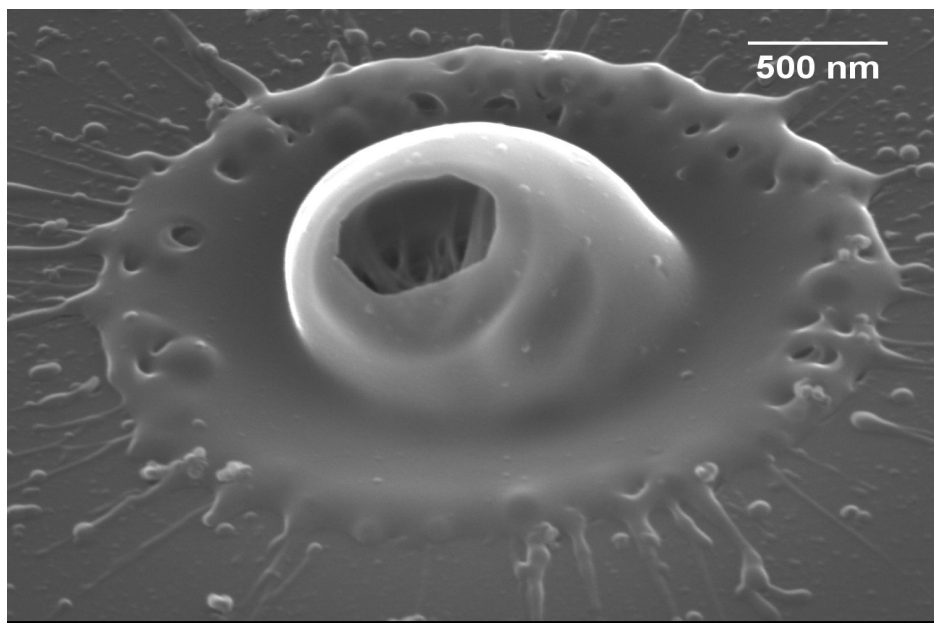

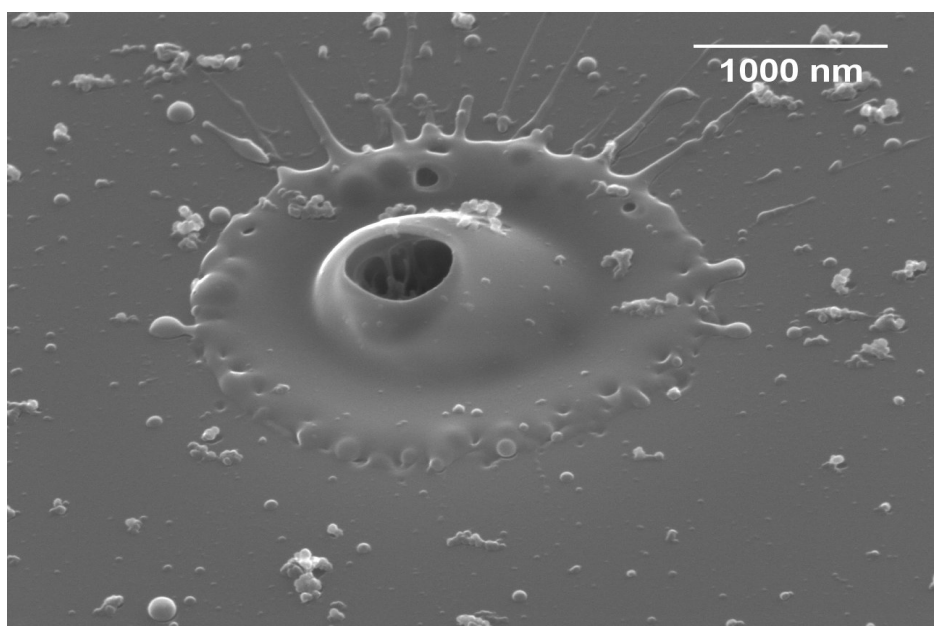

**Figure 17:** FESEM micrograph of the surface modified region at fluence (a) 4.6 J/cm$^2$, (b) 4.24 J/cm$^2$, (a) is imaged at a magnification of 35000 x and (b) at 25000 x. Both the figures show the laser processed polymer like strands and nanopores formation inside the bumps.

# Section-F
# (Optical Surface Profiler)

This section presents the schematic depicting the fundamental working principle for the optical surface profiler (OSP) used in the study. The instrument was utilized to obtain the micrographs present in Figure 3 of the main text and data for figure 2 and supplementary section A. The detailed description of the phase shift interferometry (PSI) mode of the instrument employed in the current study can be found in *'Little, D. J. & Kane, D. M. Measuring nanoparticle size using optical surface profilers. Opt. Express **21**, 15664 (2013)'*.

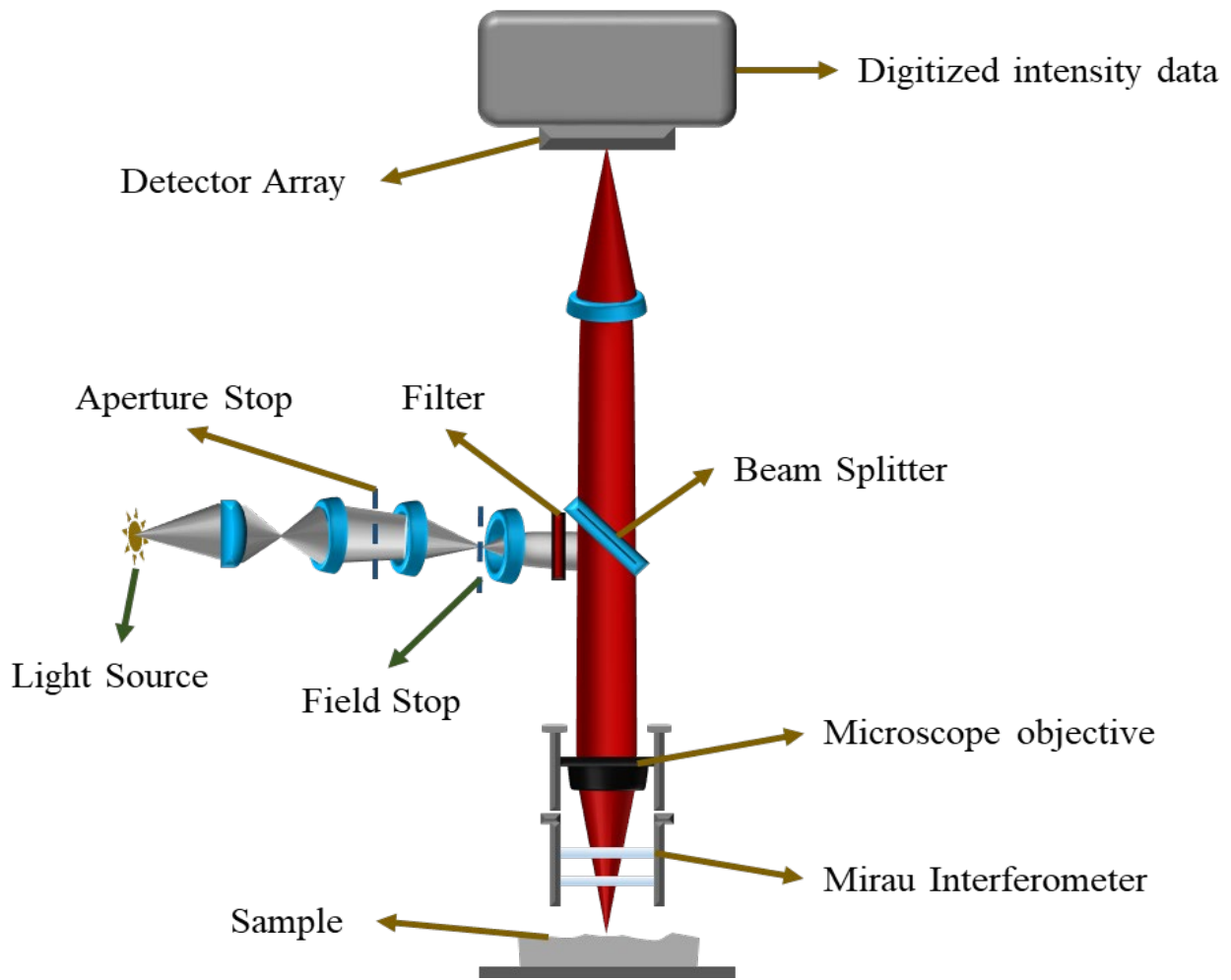

**Figure 18:** Schematic of the optical surface profiler. In the current set of studies, OSP was operated in phase shift interferometry (PSI) mode while the system is also capable to operate in the vertical shift interferometry (VSI) mode. In PSI mode, a green LED light source, centre wavelength 535 nm, bandwidth ~40 nm, is used to image the sample. Interference occurs between the light reflected from the reference mirror and that reflected from the sample to produce the interference fringes of alternating bright and dark bands. The piezoelectric transducer or PZT introduces a movement of known amount of the reference mirror to introduce a phase shift between the reference and sample beams. Intensities of the interference patterns at several relative phase shifts are analysed computationally to obtain the 3D profile. In depth operational and mathematical analysis of the OSP is discussed in, '*Little, D. J. & Kane, D. M. Measuring nanoparticle size using optical surface profilers.* Opt. Express **21**, 15664 (2013).'.